**Advance Real-time Detection of Traffic Incidents in Highways using Vehicle Trajectory Data.**

**Sudipta Roy**
Department of Civil, Environmental, and Construction Engineering
University of Central Florida,
Orlando, Florida, United States, 32816
Email: sudipta.roy@ucf.edu
ORCID: 0000-0002-9588-6013

**Samiul Hasan**
Department of Civil, Environmental, and Construction Engineering
University of Central Florida, Orlando, Florida,
United States, 32816
Email: samiul.hasan@ucf.edu
ORCID: 0000-0002-5828-3352

*Word Count: 4,957 words + 4 tables (250 words per table) = 5,957 words*

*Submitted: August 01, 2024*

**ABSTRACT**

A significant number of traffic crashes are secondary crashes that occur because of an earlier incident on the road. Thus, early detection of traffic incidents is crucial for road users from safety perspectives with a potential to reduce the risk of secondary crashes. The wide availability of GPS devices now-a-days gives an opportunity of tracking and recording vehicle trajectories. The objective of this study is to use vehicle trajectory data for advance real-time detection of traffic incidents on highways using machine learning-based algorithms. The study uses three days of unevenly sequenced vehicle trajectory data and traffic incident data on I-10, one of the most crash-prone highways in Louisiana. Vehicle trajectories are converted to trajectories based on virtual detector locations to maintain spatial uniformity as well as to generate historical traffic data for machine learning algorithms. Trips matched with traffic incidents on the way are separated and along with other trips with similar spatial attributes are used to build a database for modeling. Multiple machine learning algorithms such as Logistic Regression, Random Forest, Extreme Gradient Boost, and Artificial Neural Network models are used to detect a trajectory that is likely to face an incident in the downstream road section. Results suggest that the Random Forest model achieves the best performance for predicting an incident with reasonable recall value and discrimination capability.

**INTRODUCTION**

Traffic crashes significantly impact various aspects of daily life, ranging from time loss due to congestion or road blockages to physical injuries and fatalities. National Highway Traffic Safety Administration (NHTSA) estimated that, approximately 40,990 individuals lost their lives in traffic crashes in the United States in 2023 (*1*). A considerable number of traffic crashes are secondary crashes that occur because of an earlier incident on the road. These secondary crashes are particularly hazardous and contribute to the overall frequency and severity of traffic crashes. Previous studies reported that among 6.7 million crashes in 2019, approximately 9.2% of those crashes were caused by a preceding incident (2, 3). Detecting a road incident early is essential for warning drivers about the potential of a secondary crash. An effective road incident detection system or secondary crash alert system can significantly reduce crash frequency and severity and help alleviate congestion, thereby saving drivers both time and money (4). With increasing traffic crashes and congestion on roadways, the need for a real-time incident detection system has never been more important.

Previous research has extensively investigated how to predict crash risks in real time (5). Typically, crashes are predicted in real time at the level of a road segment using detector data aggregated over a time interval (e.g., 5 min). There are some limitations of using detector data due to low coverage or inconsistent/missing information. Detector data also lacks individual driving attributes which are key to identify the secondary crash potential (6–9). The goals of real-time crash risk prediction is to assist traffic managers to mitigate the traffic impacts of a crash as well as to alert drivers about the incident to reduce the risk of any secondary crash. However, such predictions have limits in preventing a secondary crash event as those predictions are made based on aggregated detector data, leading to latency in communicating the risk to individual drivers. An alternative to aggregate level crash risk prediction is to detect preceding incident in real time for predicting the risk of a secondary crash for individual drivers.

The widespread adoption of the Global Positioning System (GPS) has created an opportunity to use connected vehicle trajectory data or floating car data (FCD) for real-time incident detection to overcome the limitations of detector data. These trajectory data offer valuable information such as vehicle speed, travel times, as well as details regarding trip origins, destinations, chosen routes, and the traffic state with much more wider coverage area in significantly lower cost (10). This data allows for comprehensive coverage, real-time monitoring, and detailed insights into traffic patterns and driver behaviors, which improve the accuracy of incident detection. Consequently, FCD is increasingly used for real-time incident detection at both aggregate and disaggregate level, offering a more efficient and cost-effective alternative to traditional methods.

This study aims to investigate how real-time vehicle trajectory data can be used to advance detect a traffic incident occurred in the downstream segment of a vehicle. Thus, the objective of this study is to identify traffic incidents from real-time individual connected vehicle trajectory data using different machine learning algorithms. This study holds significant potential for both drivers and traffic management agencies. The early detection of traffic events will be beneficial for drivers by alerting them about any potential secondary crash. It will also impact traffic incident management system by developing a response plan early for reducing congestion and improving overall traffic flow.

**LITERATURE REVIEW**

Traditionally, researchers use traffic detector/loop detector data for traffic incident detection and risk prediction (11). However, there are several limitations of the aggregated detector level data including the inconsistency, missing data issue, and limited coverage area (7–9). Additionally, individual driver-based dynamic attributes, e.g., speed, acceleration, deceleration, lane changing information etc., can't be collected through traffic detector data (6). Vehicle trajectory data or FCD are useful to incorporate these attributes in traffic event detection models (6).

Various machine learning models have been developed for crash risk prediction and incident detections as they show superior performances by learning from data and recognizing pattern without any predefined process. The most common machine learning models include Random Forest (RF) (12–16), k-Nearest Neighbour (KNN), Extreme Gradient Boost (XGB) (17), Support Vector Machine (SVM) (13, 16, 18, 19) , and Artificial Neural Network (ANN) (13, 20). Some studies developed deep learning models for incident detection.

If trajectory data is used, anomalous trajectory detection is important for incident detection. Zualkernan et al. (21) combined Discrete Wavelet Transformation, Hidden Markov Model, and Dynamic Time Wrapping algorithms and simulated mobile phone accelerometer to train the model for vehicle incident prediction. Ma et al. (22) used Recurrent Neural Network (RNN) based autoencoders to analyze trajectory matrix similarities and detect anomalies. Yang et al. (23) used trajectories from multiple vehicles to detect incident with very high recall value. Some of these studies used real-world trajectory data and the rest used simulated data. The choice of machine learning (ML) or deep learning (DL) models will depend on data type, quantity, and purpose of study. **Table 1** presents some relevant studies for detecting traffic incidents using ML and DL-based models.

**Table 1. Detection of traffic incidents using machine learning or deep learning models**

| Study | Model used | Parameters used | Data Source |
|---|---|---|---|
| Yuan et al., 2003 (18) | SVM | Volume, occupancy | Detector (Simulation) |
| Dogru et al., 2018 (13) | RF, SVM, ANN | Speed, positions of vehicle | Vehicle trajectory (Simulation) |
| Ahuja et al., 2018(15) | RF | Speed, volume, occupancy | Detector (Real world) |
| Parsa et al., 2019 (17) | XGB | Speed, volume, weather, socio-demographic data | Detector (Real world) |
| Bharath Kumar et al., 2021 (16) | RF, SVM, ANN | Speed, positions of vehicle | Vehicle trajectory (Simulation) |
| Xie et al., 2022 (14) | RF, KNN | Flow, speed, occupancy | Detector (Real world) |
| Elsahly et al., 2023 (12) | RF | Flow rate, detector occupancy, distance between detectors | Detector (Simulation) |
| Koetsier et al., 2022 (19) | SVM | Relative position, speed, acceleration | Vehicle trajectory (Real world) |
| Zyryanov et al., 2023 (20) | ANN | Volume, speed | Detector (Simulation) |
| Yang et al.., 2021 (24) | DCNN | Speed, position, acceleration, deceleration, heading angle etc. | Vehicle Trajectory (Simulator) |
| Zualkernan et al., 2018 (21) | DWT-HMM-DTW | Acceleration mean, std, max, entropy, energy | Vehicle Trajectory (Real world simulation) |
| Wang et al., 2024 (25) | LAGMM | Vehicle Speed, speed of platoon, steering wheel angle | Vehicle Trajectory (Real world) |
| Ma et al., 2019(22) | RNN | position | Vehicle Trajectory (Real world) |

One of the key issues of incident detection models is the imbalanced data issue. Previous studies have addressed this issue by oversampling the incident data. Peng et al. (26) found that 1:4 ratio of incident and non-incident data is the best for the classification problems which is hardly available for this type of modeling. To resolve this issue, studies (14, 27) used Synthetic Minority Oversampling Technique (SMOTE) and Self-adaptive Synthetic Over-sampling technique (SASYNO) to over-sample the severely

*Roy and Hasan*

imbalanced data to detect incidents. Huang and Chen et al.(28) used both under-sampling with Repeated Edited Nearest Neighbors (RENN) and over-sampling with SMOTE to solve the imbalance issue. Another key issue for imbalanced data classification problem is the determination of classification threshold. Peng et al. (26) found the Youden index method best to set the classification threshold. Yang et al. (29) found the minimum cross-entropy method as the best suitable method for this issue.

To the best of our knowledge there has been no work focusing on predicting an incident ahead of certain downstream distance in a highway. Additionally, there is a gap in understanding how to effectively use trajectory data with irregular intervals for this purpose. This study aims to address these gaps by utilizing vehicle trajectory data for real-time downstream road incident detection using machine learning algorithms. This study has two contributions. *First*, we develop a highway traffic incident detection framework that leverages limited length connected vehicle trajectory data. *Second*, we are using low-frequency trajectory data here with irregular interval which could pave the way for utilizing this more cost-effective data option in traffic incident detection modeling. This work will significantly enhance driving safety by helping the drivers to avoid secondary crashes and will assist the traffic incident management authorities in implementing efficient response plans to mitigate congestion impacts.

## DATA DESCRIPTION AND EXPLORATION

### Study Area

The study area of this analysis encompasses a 273-mile section of the I-10 interstate highway in Louisiana, extending from St. Tammany parish in the east to Calcasieu parish in the west (**Figure 1**). It links 15 Coastal Parishes and some major cities of Louisiana like New Orleans, Baton Rouge, and Lafayette to the south, south-eastern and south-western states of USA. This section of I-10 is one of the most crash-prone highways in Louisiana having 54 fatalities alone in 2021 (30). Different recurrent and non-recurrent incidents result in frequent congestions in this interstate highway (31, 32).

### Connected Vehicle Trajectory Data

Connected vehicle trajectory data is collected from Otonomo for three consecutive days (August 27 to 29 in 2021) in Louisiana state during the evacuation period of Hurricane Ida. The mandatory evacuation order for Ida was placed two days before the hurricane landfall (August 29, 2021). It is expected that during the evacuation period, as one of the major highways of Louisiana, I-10 would experience a high volume of traffic (27). This high traffic volume during evacuation would increase the likelihood of a crash occurrence (33). The raw data contains the trajectory information of **135,204** individual vehicles. For each trajectory point, the coordinates of the point, the name of the roads and nearby locations as well as the mobility statistics including speed and mobility angle of the vehicles (travel direction with respect to north) are provided. The frequency of the trajectory data is varying and less than 30 seconds.

### Traffic Incident Data

The incident dataset for the study period (27th August 2021 to 29th August 2021) is collected through Regional Integrated Transportation Information System (RITIS) database (34). This study considers only incidents that resulted in the closure of one or more travel lanes and there are **256** such events for I-10 corridor in Louisiana during the study period (**Figure 1**).

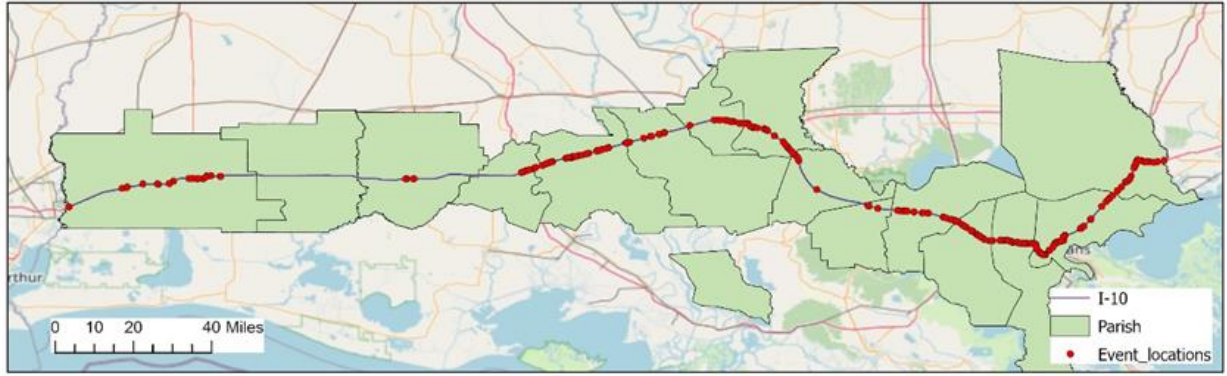

**Figure 1. Study area (I-10 in Louisiana) and Event locations with incident occurrence**

**Figure 2** illustrates several key aspects of the incidents. The top-left histogram shows that the majority of incidents are cleared within 15 minutes and few incidents last longer. The top-right bar chart reveals that stalled vehicles are the most common type of incident (200 cases), followed by accidents (48 cases). The bottom-left chart highlights that more incidents occur in the westbound direction of I-10 (153 cases) compared to the eastbound direction (103 cases). Lastly, the bottom-right chart indicates only a small number involving the closure of two lanes (14 cases). Due to low sample size of incidents in the data, we do not consider the variation in incident types and number of lanes closed.

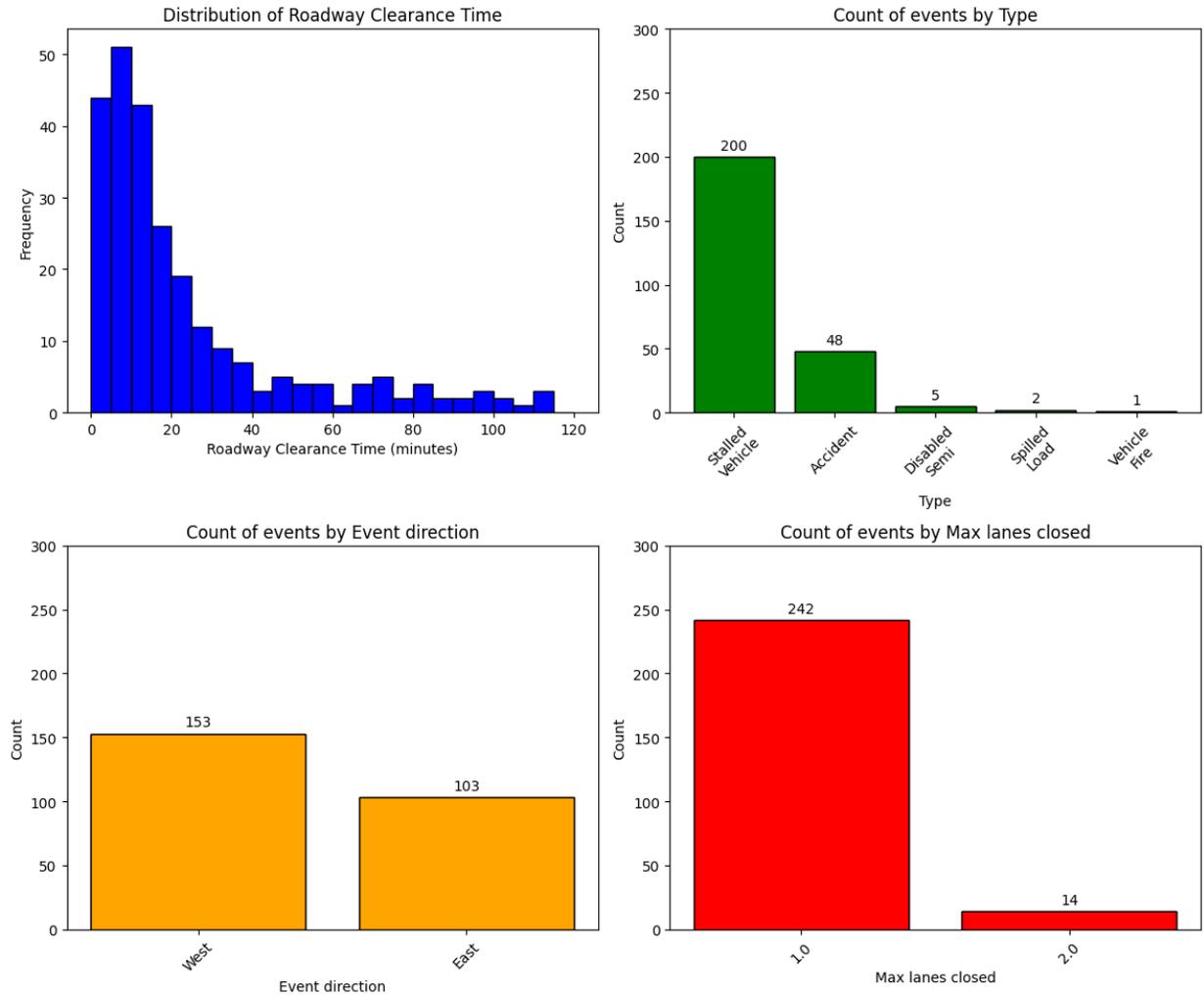

**Figure 2. Different attributes of the traffic incidents**

### DATA PREPARATION METHODS
This section includes detailed procedure of trajectory data processing, trajectory database building, trajectory-event data spatial-temporal matching and weather information processing.

**Trajectory Data Processing**
 The raw trip trajectory data needs to undergo filtering and trip segregation process as a part of data preparation to get the trip information on I-10. The entire trajectory data processing is shown in **Figure 3**. The inputs are in parallelogram, start/end of process are in oval, decision-making steps are in diamond shapes and the operation steps are in rectangle shape.

- First, the vehicle trajectories which did not contain any point coordinates on I-10 are removed from dataset. It reduces the number of vehicle trajectories to **31,182** for three days.
- Later, the trajectory points for individual vehicles are separated to break down into individual trips. For this segregation process, two rules are followed. First it is checked if the point is on I-10 or not. The points outside I-10 are discarded. After that, the time differences between the points are considered to check if a new trip is started or not. When the time difference between two successive points is greater than 15 minutes, then the trip is terminated, and a new trip is started. After this segregation process, the total number of trip trajectories are **11,674**.

- The directions of the trips are approximated from their angle of mobility. Total number of westbound trips are **7,438** and eastbound trips are **4,236**.

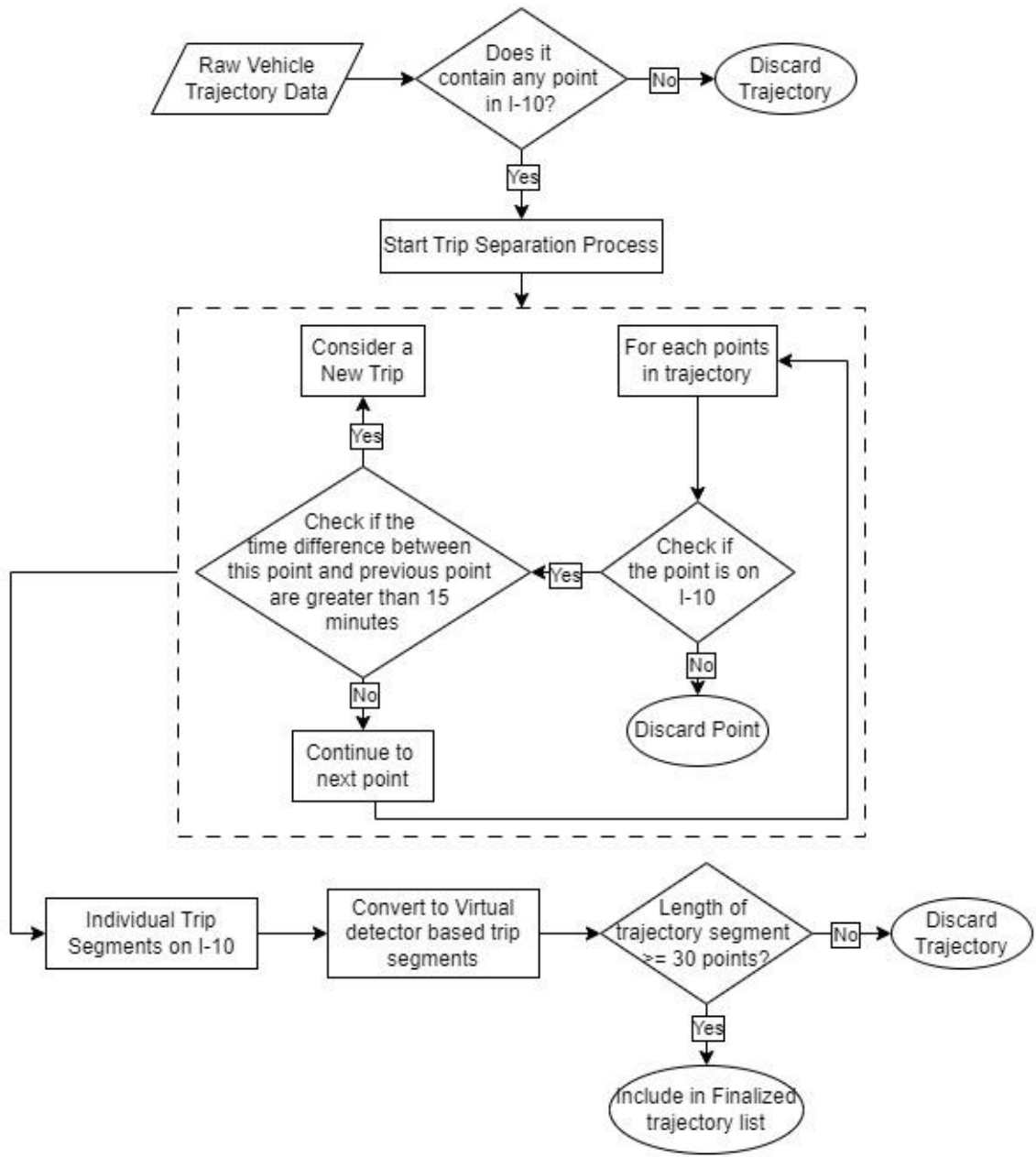

**Figure 3. Trip trajectory data processing**

- To maintain the spatial uniformity of the data, the trajectory of each trip is needed to be converted into small uniform segments. For that purpose, a set of virtual detectors are considered on both I-10 eastbound and I-10 westbound directions with 110 yards (1/16 mile) gap between successive detectors. The travel time, mobility angle and mobility speed of the vehicle on each trip are interpolated on the detector positions. As a result, the trajectories are converted to detector position-based trajectories (**Figure 4**).
- Weather data is extracted to check if any rain was present during the trip (35).

*Roy and Hasan*

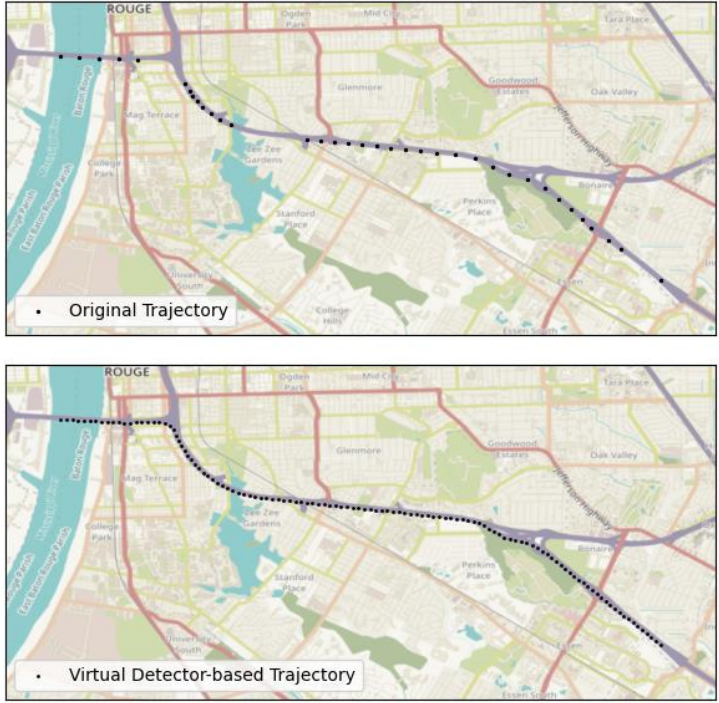

**Figure 4. Original and detector location-based trajectory example**

**Building a Detector Database**

    A database based on virtual detectors is developed with the converted trajectories. The variable list and exploratory information of the detector database for both directions are shown in **Table 2**. Peak periods for the road use are assumed to be from 6 AM to 10 AM and 3 PM to 7 PM and the rest of the day is assumed as off-peak period (36). This detector database will be used in the modeling later.

**TABLE 2. Detector database exploratory information**

| Eastbound Detectors | | | | | |
|---|---|---|---|---|---|
| **Features** | unit | mean | std | min | max |
| Detector id | Total number = 4201 | | | | |
| Number of peak period observations | - | 80 | 49 | 23 | 227 |
| Peak mean speed | mph | 68.7 | 9.3 | 32.3 | 79.5 |
| Peak speed standard deviation | mph | 6.1 | 4.5 | 1.9 | 25.5 |
| Number of off-peak period observations | - | 100 | 48 | 27 | 235 |
| Off-peak mean speed | mph | 70.0 | 6.7 | 42.9 | 77.7 |
| Off-peak speed standard deviation | mph | 5.7 | 4.2 | 2.5 | 23.6 |
| Mobility heading angle | - | 87.1 | 26.3 | 4.0 | 171.0 |
| **Westbound Detectors** | | | | | |
| Detector id | Total number = 4210 | | | | |
| Number of peak period observations | - | 151 | 43 | 11 | 243 |
| Peak mean speed | mph | 62.4 | 9.0 | 25.5 | 79.5 |
| Peak speed standard deviation | mph | 11.3 | 5.7 | 3.1 | 27.3 |
| Number of off-peak period observations | - | 203 | 62 | 14 | 330 |
| Off-peak mean speed | mph | 62.4 | 7.7 | 36.0 | 78.9 |
| Off-peak speed standard deviation | mph | 10.6 | 5.5 | 3.1 | 29.8 |
| Mobility heading angle | - | 266.6 | 26.5 | 175.0 | 351.0 |

For both periods, the mean and standard deviation of travel speed and heading angle information is collected through the previously processed detector location-based trajectory data. The trip counts and mean speeds in the eastbound and westbound road detectors are shown in **Figure 5**. The visualization suggests for both peak and off-peak periods, there are enough connected vehicle counts to build the virtual detector-based historical traffic database. Since the data was collected during an evacuation period, there is minimal variation between the peak and off-peak traffic count profiles. Consequently, the mean speeds for both peak and off-peak periods are very similar, which is atypical during regular periods. Nevertheless, the historical traffic database retains the distinctions between peak and off-peak periods for generalization purposes.

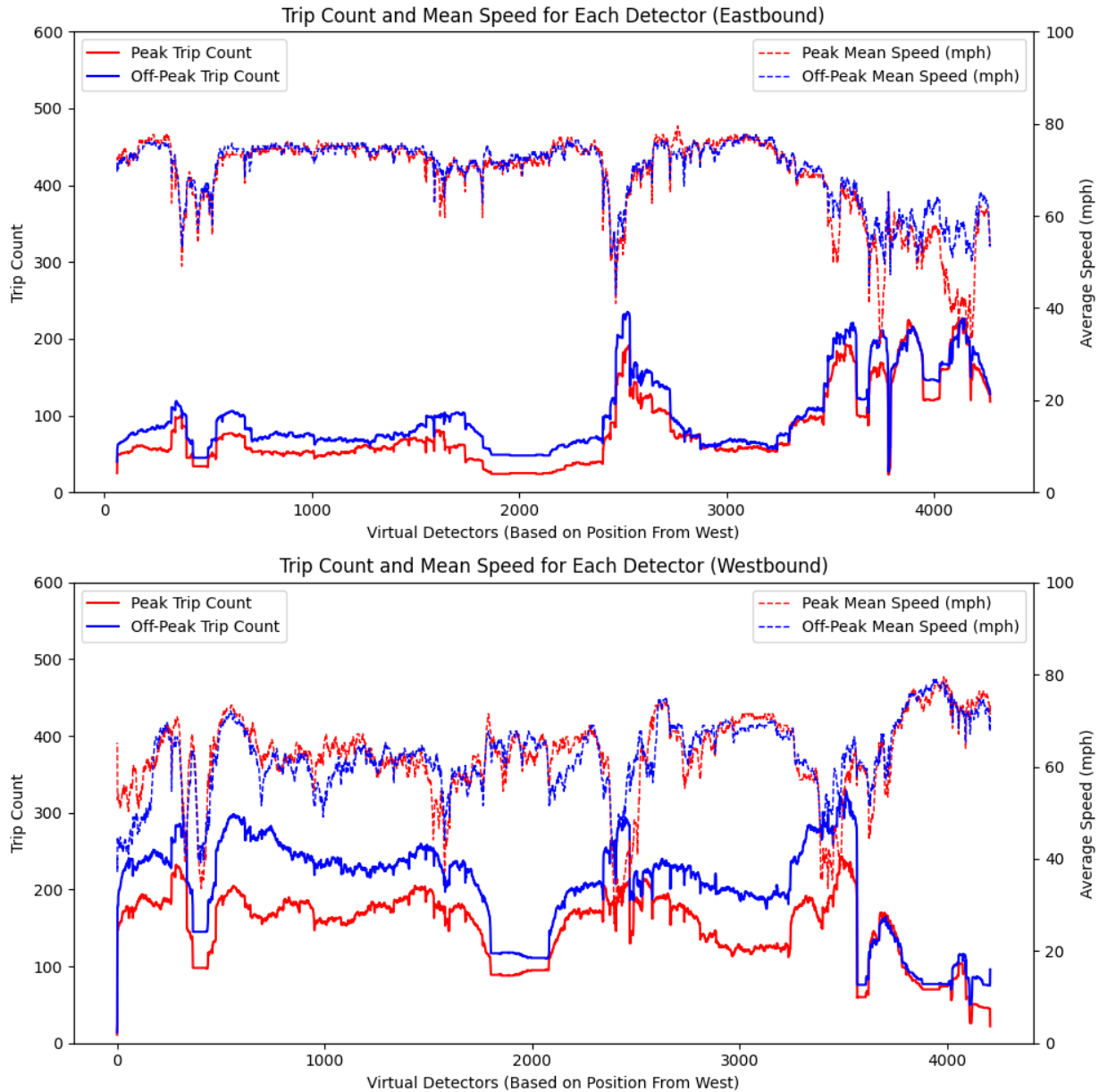

**Figure 5. Trip counts and mean speeds on virtual detectors.**

**Spatial-Temporal Matching of Traffic Event and Trajectory Data**
The traffic incidents and the trip trajectories are matched both spatially and temporally to find the trips which faced one or more traffic incidents during their journey period. This matching is done in several steps:
1. For each incident, the nearest virtual detector is matched which is referred here as event detector.
2. The detector location-based trip trajectories are examined for each event to determine whether the event detector lies within the trajectory.
3. Then it is checked if the trajectory passes the event detector within 2 hours before the event start time or within 15 minutes after the event start time. If the trajectory coinciding time with the event location is within 2 hours before the event start time, it is considered normal. Otherwise, if the trajectory coinciding time with the event location is within 15 minutes of *event start time*, it is assumed the trajectory is affected for this event. If the trajectory doesn't pass through the event detector within the above timeframe, it is discarded. **Figure 6** shows the spatial-temporal matching process.

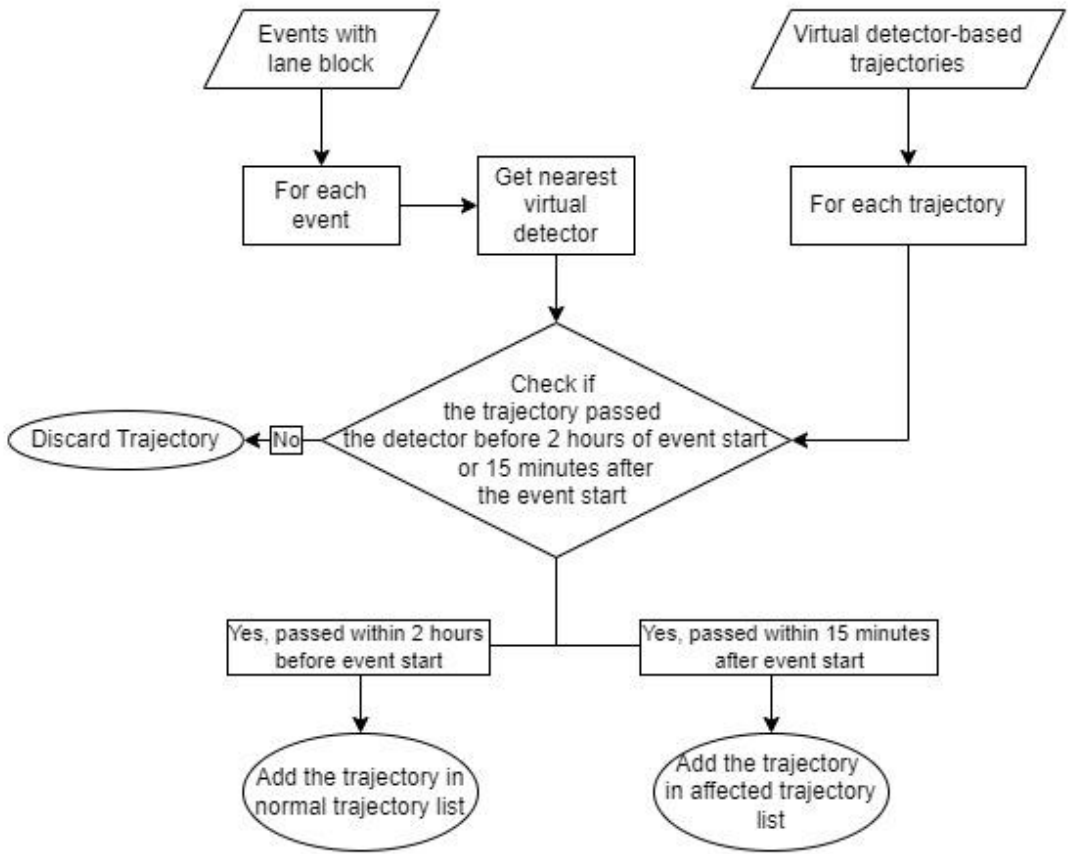

**Figure 6. Event-trajectory spatial-temporal matching process.**

- **Figures 7(a) and 7(b)** demonstrate two representative event scenarios for comparing the speed of affected trajectories against the speed of normal trajectories. It is seen that for both cases there is a significant drop in speed for the affected trajectories. However, the normal trajectories are not demonstrating such drop.
- The trajectories where at least 16 detectors or 1760 yards (1 miles) are available on the upstream of the event detector are included in the final modeling dataset. Only these 16 detector locations data are considered for analysis.

The matching result yields that 342 trip trajectories faced at least one event with lane closing occurrence on their way within 15 minutes of that event occurrence. We considered another 2265 trajectories which pass the event detector within 2 hours before the event occurrence that means they are unaffected or normal. So, the total number of trajectories used for modeling here is 2607.

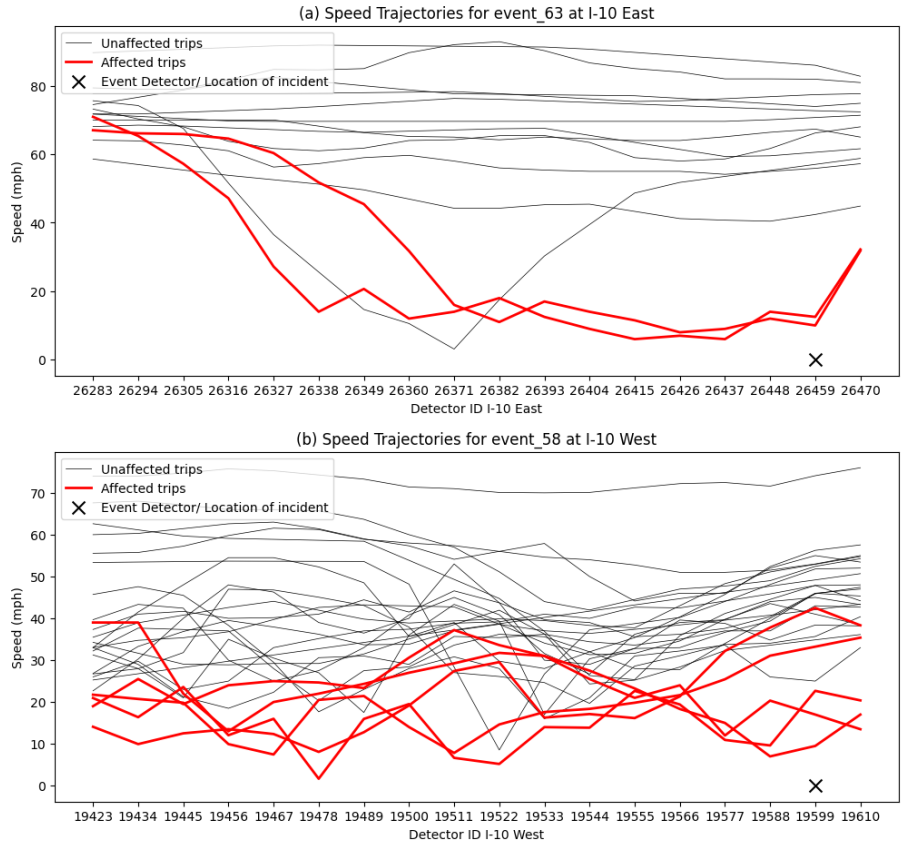

**Figure 7. Visualization of affected and unaffected trajectories due to incidents**

## TRAJECTORY CLASSFICATION METHOD
This section includes target feature, selection of algorithms for modeling, balancing the dataset through applying oversampling techniques, and performance metrics.

### Target Feature
In this study, we formulate a classification problem where the target feature will be for every individual trip trajectory whether any travel lane-blocking incident happened or not in 440 yards (0.25 miles) downstream of its current position within the last 15 minutes. **Figure 8** provides a conceptual diagram for it.

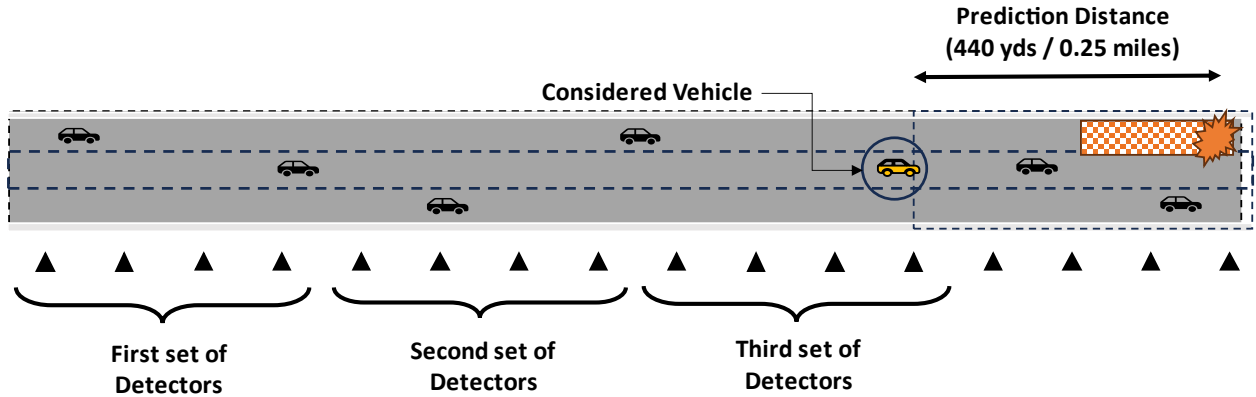

**Figure 8. Conceptual diagram for advance downstream road incident detection**

**Feature Extraction for Modeling**

Different available features are extracted from the selected trip trajectories which are mainly related to the speed of the vehicles in different points of the trajectory and the historical trajectory data (**Table 3**). The detectors on the upstream of the event are grouped into 3 sets (**Figure 8**). Each set contains 4 consecutive detectors and 3 features are collected for each detector set: the mean speed, the standard deviation of speed and the historical mean speed on those detectors. Two dummy variables are also added: the rain indicator and the peak/off-peak period indicator.

**TABLE 3. Feature exploration for modeling**

| Features | Unit | mean | std | min | max |
|---|---|---|---|---|---|
| Mean speed on first set of detectors | mph | 51.5 | 22.7 | 2.8 | 100.8 |
| Standard deviation of speed on first set of detectors | mph | 2.9 | 3.2 | 0.0 | 22.0 |
| Historical mean speed on first set of detectors | mph | 58.7 | 9.2 | 29.6 | 77.1 |
| Change in heading angle on first set of detectors | degree | 3.8 | 8.76 | 0.0 | 92.6 |
| Mean speed on second set of detectors | mph | 51.0 | 22.9 | 1.9 | 96.5 |
| Standard deviation of speed on second set of detectors | mph | 3.0 | 3.3 | 0.0 | 29.6 |
| Historical mean speed on second set of detectors | mph | 58.7 | 9.2 | 28.0 | 76.9 |
| Change in heading angle on second set of detectors | degree | 4.2 | 9.5 | 0.0 | 90.3 |
| Mean speed on third set of detectors | mph | 50.3 | 23.0 | 2.0 | 97.3 |
| Standard deviation of speed on third set of detectors | mph | 3.0 | 3.3 | 0.0 | 34.1 |
| Historical mean speed on third set of detectors | mph | 58.4 | 9.1 | 29.6 | 77.3 |
| Change in heading angle on third set of detectors | degree | 4.2 | 8.1 | 0.0 | 70.9 |
| Peak | Yes: 1125, No: 1482 | | | | |
| Rain | Yes: 288, No: 2319 | | | | |
| Affected trajectory? (Target) | Yes: 342, No: 2265 | | | | |

**Applying data balancing technique**

Since the number of affected trajectories is very low compared to the total number of trajectories (approximately 1:7 ratio), following previous studies (27, 28), we use **Synthetic Minority Oversampling Technique (SMOTE)** on the training data to improve the ratio. The ratios used in SMOTE method are 0.25, 0.5, and 1.0.

**Algorithms**

As this is a classification problem related to road event detection, Logistic Regression (LR) is a commonly used approach in this type of problems. We will also use three other state-of-the-art machine learning-based algorithms in this study: Random Forest (RF), Extreme Gradient Boosting (XGB) and Artificial Neural Network (ANN). We have done 5-folded cross validation here. The parameters of each model are tuned using *grid search* algorithm to find best output from the models (**Table 4**). As we have different sets of data here (after data balancing using SMOTE), we separately tuned the model parameters for all sets of data.

**TABLE 4. Tuned parameters for different models**

| Model | Parameters |
|---|---|
| Logistic Regression (LR) | Regularization strength, penalty, solver |
| Random Forest (RF) | number of trees, maximum depth of tree |
| Extreme Gradient Boosting (XGB) | learning rate, maximum depth of tree, number of trees, subsample ratio |
| Artificial Neural Network (ANN) | activation function, alpha (strength for L2 regularization), batch size, learning rate |

**Performance Metrics**

The recall, false alarm rate, and area under the ROC curve parameters are analyzed from the model outputs are considered to evaluate the model performance (37).

*Recall*

Recall or sensitivity or true positive rate measures the proportion of actual positives or in this modeling context, the correctly predicted affected trajectories divided by the number of affected trajectories.

$$Recall = \frac{Correctly\ predicted\ affected\ trajectories}{Total\ number\ of\ affected\ trajectories}$$

*False alarm rate (FAR)*

False alarm rate or false positive rate measures the proportion of false positive instances or in this modeling context, the incorrectly predicted affected trajectories divided by the number of normal trajectories.

$$False\ alarm\ rate = \frac{Incorrectly\ predicted\ affected\ trajectories}{Total\ number\ of\ normal\ trajectories}$$

*Area under the receiver operating characteristics curve (AUC-ROC)*

AUC-ROC summarizes the performance of the classifier across different threshold values. The receiver operating characteristics (ROC) curve plots recall against false alarm rate for different threshold values. AUC-ROC range varies from 0 to 1, where 0.5 implies a classifier with no discriminative capabilities and 1 implies a perfect classifier.

$$AUC - ROC = \int_0^1 Recall * d(False\ Alarm\ Rate)$$

**Classification threshold set**

We have set the optimum classification threshold using Youden's Index (38). This is calculated using ROC curve where recall vs. FAR are plotted for different threshold value. Now,

$$Youden's\ Index = Recall - False\ Alarm\ Rate$$

For the maximum Youden's Index value, the corresponding classification threshold is chosen as optimum threshold. The optimum threshold is recalculated for each method, SMOTE ratio, and iteration.

## RESULTS

**Model Outputs**
  **Figure 9** shows the mean prediction outputs from 5-folded of cross validation across different applied algorithms and different SMOTE ratio. The LR model is performing the worst for this prediction purpose. With the original condition of dataset (without oversampling) or after oversampling both the recall and FAR are zero and the AUC-ROC is 0.50, which means the model can't identify any event, whether it is true or false positive and there is no discriminative capability of the model at all.

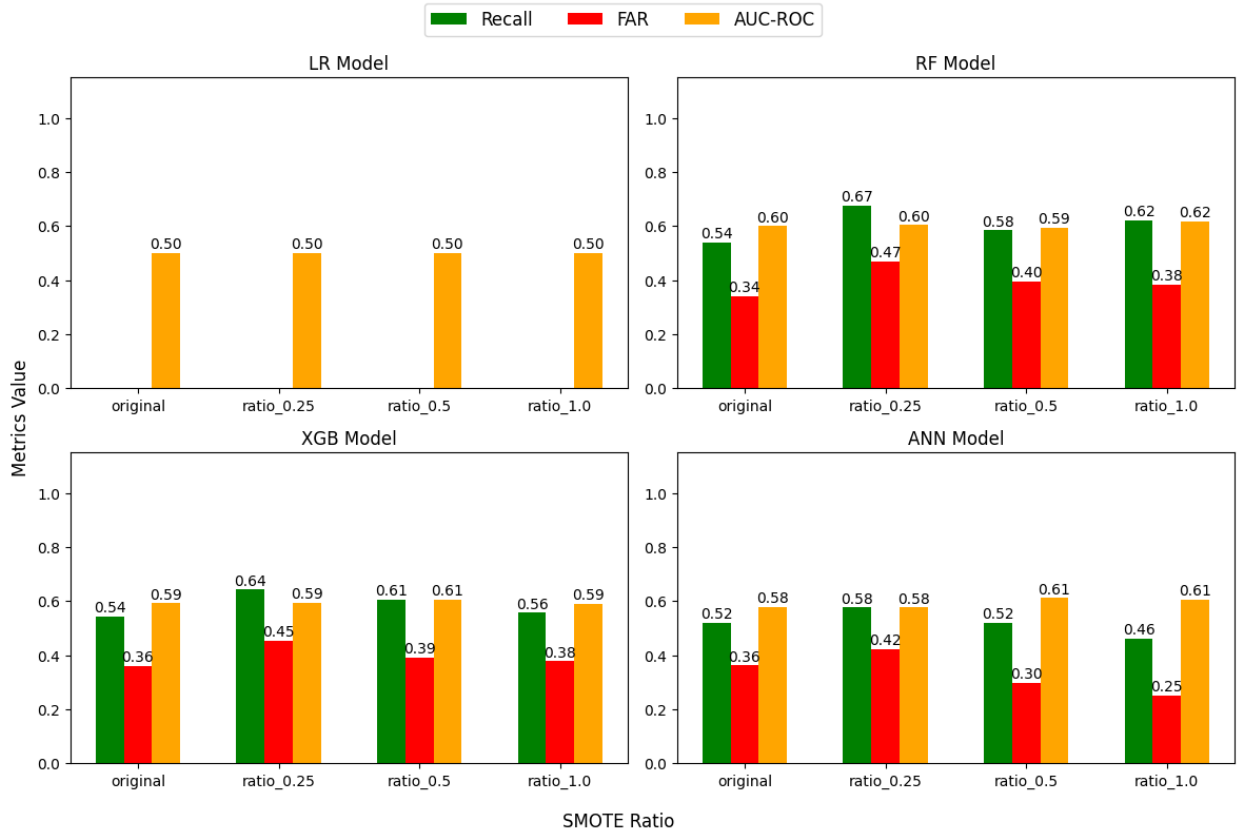

**Figure 9. Model Outputs**

  The performances of RF, XGB, and ANN models are almost similar. For all the models, the mean recall value is highest For SMOTE ratio 0.25, but it is also associated with very high FAR. The mean AUC-ROC value stabilizes to approximately 0.60 before and after applying different SMOTE ratio, and this indicates there are limited discriminative capabilities developed on the trained models to identify any downstream potential incident. The highest mean recall value (0.67) is obtained for the RF model, which indicates 67% times this model can identify any incident in case there are any real incident occurrence. The FAR value (0.47) in this model is also highest indicating this model will predict 47% of the normal trajectories incorrectly as incident affected. Overall, the RF model for SMOTE ratio 1.00 performs best as the mean AUC-ROC is highest for this model.

**Feature Importance**
  The feature importance here is analyzed from the RF model with SMOTE ratio 1.00 and sorted by feature importance in **Figure 10**. The analysis suggests the change in heading angle in the third detector set and the mean speed on the third detector set, which contains the immediate position of the vehicle, are two most important to predict any downstream accident. Another aspect of the analysis is the absence of any exceptionally important feature in the model.

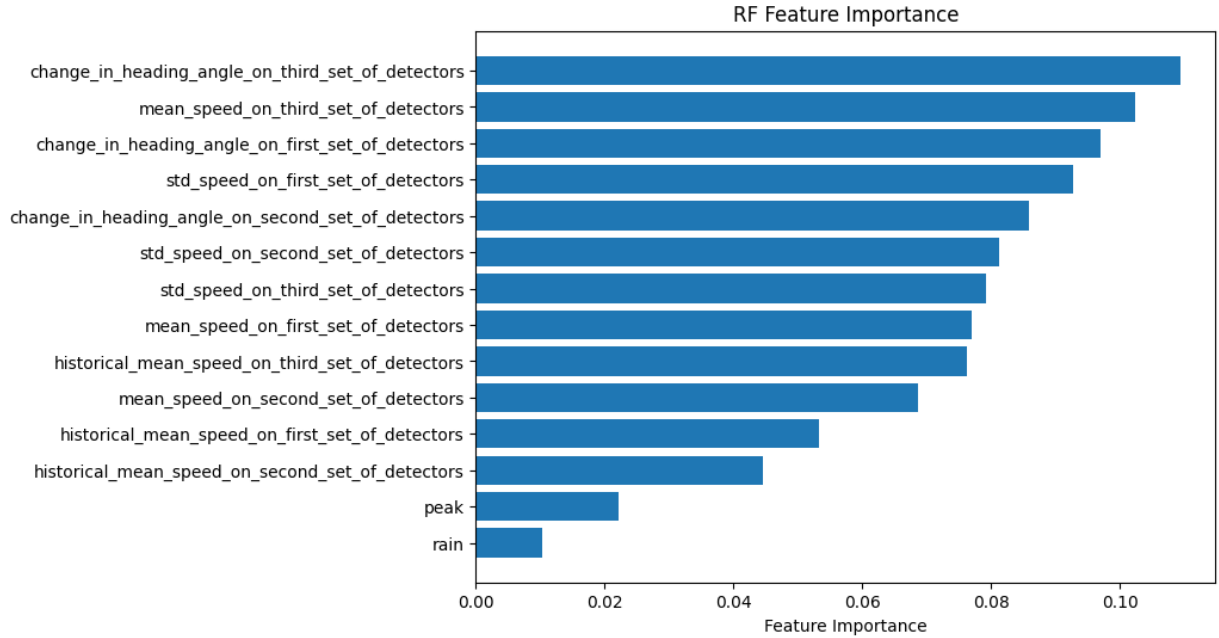

**Figure 10. Feature Importance**

**CONCLUSIONS**

Advance detection of downstream road incidents is significant for enhancing driver's safety as it provides drivers an opportunity to become cautious and reduce the chance of any secondary crash. In this study, a data-driven methodological framework has been developed to convert trip trajectories of unequal timesteps obtained from connected vehicles to trajectories based on virtual detectors. Then using machine learning algorithms, any downstream incident is predicted from the trajectory information available to the vehicle.

This research has a significant potential to improve traffic safety by contributing to the zero deaths vision of the U.S. (39). This work has developed a framework for real-time traffic event predictions incorporating short-term trajectory data and historical traffic data. Accurate and reliable predictions of traffic incidents will alert drivers for any unexpected incident in downstream and reduce the chance of a secondary crash. It will also benefit traffic management systems to reduce non-recurrent congestion with faster response plans.

The study has some limitations that can be addressed in future. For instance, the RITIS event data does not contain enough information of the events such as how long the lanes were closed for an event, how was the event intensity, if it was impacting the traffic flow or not, and if any road shoulder was temporarily used or not. Perhaps some event dataset with this kind of information will make the framework more precise. As the trajectory data provided by Otonomo has unequal timesteps and numerous missing values, we had to do extensive data processing to make the trajectories generalized to extract modeling data, in which effort we might also lose some valuable insight.

**ACKNOWLEDGEMENT**
The research presented in this paper was supported by NSF EAGER Grant #2122135. However, the authors are solely responsible for the findings presented here.
ChatGPT was used when preparing the draft of this manuscript to assist in reconstructing or polishing some author-provided text only.

**AUTHOR CONTRIBUTIONS**
The authors confirm contribution to the paper as follows: study conception and design: Roy, Hasan; data collection: Hasan; analysis and interpretation of results: Roy, Hasan; draft manuscript preparation and editing: Roy, Hasan; funding acquisition and supervision: Hasan. All authors reviewed the results and approved the final version of the manuscript.